# Extreme-value analysis in nano-biological systems: Applications and Implications

Kumiko Hayashi[1], Nobumichi Takamatsu[1] and Shunki Takaramoto[1]

[1]Institute for Solid State Physics, The University of Tokyo, Kashiwano-ha 5-1-5, Kashiwa, Chiba 277-8581, Japan
*Correspondence to Kumiko Hayashi
hayashi@issp.u-tokyo.ac.jp

**ABSTRACT**

Extreme value analysis (EVA) is a statistical method that studies the properties of extreme values of datasets, crucial for fields like engineering, meteorology, finance, insurance, and environmental science. EVA models extreme events using distributions such as Fréchet, Weibull, or Gumbel, aiding in risk prediction and management. This review explores EVA's application to nanoscale biological systems. Traditionally, biological research focuses on average values from repeated experiments. However, EVA offers insights into molecular mechanisms by examining extreme data points. We introduce EVA's concepts with simulations and review its use in studying motor protein movements within cells, highlighting the importance of *in vivo* analysis due to the complex intracellular environment. We suggest EVA as a tool for extracting motor proteins' physical properties *in vivo* and discuss its potential in other biological systems. While there have been only a few applications of EVA to biological systems, it holds promise for uncovering hidden properties in extreme data, promoting its broader application in life sciences.

**Introduction**
    Extreme value analysis (EVA) is a branch of statistics that focuses on extreme values. It is the study of the statistical properties of particularly large or small values within a data set. EVA is widely applied in various fields where extreme phenomena hold significant importance (Coles 2001; Gilleland and Katz 2016), such as disaster prevention (de Haan and Ferreira 2006; Tippett et al. 2016), finance (Kratz 2017), safety estimation (Songchitruksa and Tarko 2006), sports (Einmahl and Magnus 2008; Gembris et al. 2002), human lifespan (Dong et al. 2016; Rootzen and Zolud 2017), and the recent pandemic (Wong and Collins 2020). Recently, its applications in the biological data analysis has also become active (Basnayake et al. 2019; Tsuduki 2024). The central concepts of EVA involve identifying the largest or smallest values in a dataset, determining whether these extremes follow distributions like the Fréchet distribution, Weibull distribution, or Gumbel distribution, and analyzing data points that exceed specific thresholds to estimate their distribution. EVA is utilized in a broad range of fields. For instance, in meteorology, it is used to predict and assess extreme weather events such as typhoons, floods, and droughts. In finance, it aids in risk management for extreme price fluctuations, such as market crashes or surges. In engineering, it is applied to analyze extreme stresses or loads for evaluating the durability and safety design of structures. EVA serves as a powerful tool for predicting the frequency and impact of extreme events, aiding in risk management and safety measures.

    In this review paper, we discuss the application of EVA to the analysis of experimental data in nanoscale biological systems. By using EVA, it is possible to explore novel molecular mechanisms in life sciences through the behaviour of extreme values in the data. Rare events in nanoscale biological systems, such as protein misfolding (Harada et al. 2015) and synaptic delay (Tsuduki 2024), have already been studied, and their prediction is considered important due to their relevance to diseases. In the following sections, we first learn the concepts of EVA through simple simulations using random numbers. Next, we introduce the application of EVA to the movement of motor proteins within cells, investigated in our recent research paper (Naoi et al. 2024). The physical properties of motor proteins, such as force and velocity, have been investigated by *in-vitro* single-molecule experiments, in which the functions of motor proteins consisting of minimal complexes were analyzed in glass chambers (Brenner et al. 2020; Elshenawy et al. 2019; Gennerich et al. 2007; Hirakawa et al. 2000; Mallik et al. 2004; Schnitzer et al. 2000; Toba et al. 2006). However, because motor proteins function fully in the intracellular environment and are equipped with accessory proteins, the investigation of motor proteins *in vivo* is as important as *in-vitro* single-molecule experiments. We believe that EVA can be useful as a new information science tool for extracting the physical properties of motor proteins in the complex intracellular environment.

    Finally, we discuss the potential applications of EVA in nanoscale biological systems, with a specific example of its application to the analysis of droplet sizes in liquid-liquid phase separation (LLPS) (Takaramoto and Inoue 2024). This application of EVA was inspired by Takaramoto and Inoue's poster at the IUPAB 2024 Congress. While research utilizing EVA in nanoscale biological systems is still limited, it is believed that undiscovered properties may lie hidden within the extreme values of experimental data.

**Simulation using random numbers**
    In this section, the analytical methods of EVA are reviewed through a simple simulation using uniform random numbers ranging from 0 to 1 and Gaussian random numbers. In the case of random numbers ranging from 0 to 1, it is obvious that the finite maximum value is 1, while Gaussian random numbers do not have a finite maximum value. In other words, we examine two random numbers with different extreme properties.

    One block of EVA is considered as a data set (*i.e.*, a block) of elements $M$, from which the largest value is selected. In meteorology, $M$ is often considered as one year, which is 365 days. The selection of the block maximum ($X_{\max}^i$) from the $i$-th block (the $i$-th data set) is repeated $n$ times to collect the block maximum

data set $\{X_{\max}^i\}$ ($i = 1, \cdots, n$). Using the block maximum data set $\{X_{\max}^i\}$, the return level plot is investigated (the plot was calculated by using the ismev and evd packages in R (R Core Team 2018)) (Fig. 1a and 1b). The two axes of the return level plot (Fig. 1a and 1b, bottom) represent the return period $r_p$ and return level $z_p$. For a given probability $p$, $r_p = -\{\log(1-p)\}^{-1}$, and $z_p$ is defined by the generalized extreme value distribution as $1 - p = G(z_p)$, where

$$G(z_p) = \exp\left[-\left\{1 + \xi\left(\frac{z_p - \mu}{\sigma}\right)\right\}^{-1/\xi}\right]. \qquad (1)$$

Note that $z_p$ represents $\hat{X}_{\max}^i$, where $\{\hat{X}_{\max}^i\}$ is the rearranged data of $\{X_{\max}^i\}$, such that $\hat{X}_{\max}^1 \leq \hat{X}_{\max}^2 \leq \cdots \leq \hat{X}_{\max}^n$. Roughly, $r_p$ is the sample number. (Note that we obtained parameters of the generalized extreme value distribution $\xi$, $\mu$, and $\sigma$ in Eq. (1) by using the ismev and evd packages in R (R Core Team 2018)). Here, the return level $z_p$ refers to the magnitude of extreme values expected to occur once within the sample size we are considering, such as 100 years, 1000 people and so on. For example, when analyzing river water level data for environmental science case, its return level plot can provide information such as the maximum water level of a flood expected once every 100 years. The distribution characteristics of the extreme value data, such as the values of location ($\mu$), scale ($\sigma$), and shape parameters ($\xi$) of the extreme value distribution (the Fréchet distribution, Weibull distribution, or Gumbel distribution) can be estimated from the behavior of the return level plot (Fig. 1c).

The shape parameters of the extreme value distribution $\xi$ is important because it characterizes the behavior of extreme values, determining whether a data set has a finite maximum value or not. We found that $\xi < 0$ for the case of uniform random numbers, which have the finite maximum value (Fig. 1a), and $\xi \sim 0$ for the case of Gaussian random numbers, which do not have the finite maximum value (Fig. 1b). The existence of a finite maximum value is related to the value of the shape parameter $\xi$. Therefore, we particularly focus on the value of $\xi$ in EVA.

Depending on the value of $\xi$, it is classified into the Weibull distribution ($\xi < 0$), Gumbel distribution ($\xi = 0$), or Fréchet distribution ($\xi > 0$) (Fig. 1c). In the case of a Weibull distribution, the finite extreme value (the maximum value) $V_{\text{ex}}$ is proved to exist and is estimated using the following equation:

$$V_{\text{ex}} = \mu - \sigma/\xi \qquad (2)$$

It has been known that the extreme values (the maximum values) of a uniform random number distribution converge to a Weibull distribution (Fig. 1a), and those of a Gaussian distribution to Gumbel distribution (Fig. 1b). In our simulations, we obtained the set of parameters $(\mu, \sigma, \xi) = (0.911 \pm 0.003\text{SE}, 0.0793 \pm 0.0034\text{SE}, -0.890 \pm 0.012\text{SE})$ for uniform random numbers ranging from 0 to 1. For the set of parameters $(\mu, \sigma, \xi)$, the quantity $\mu - \sigma/\xi$ (Eq. (2)) was calculated to be 1.00 as the maximum number.

The properties of the maximum value of the simulation examples shown in Fig. 1 are known from the beginning. However, the important point is that the extreme value distribution of any data, whose extreme properties are not known, can be described by Eq. (1), based on the mathematical theorem. In unknown complex systems like nanoscale biological systems, it is a significant advantage to know the functional form (Eq. (1)) for their extreme value distributions. In the next section, we consider the application of this strength in EVA to intracellular motor proteins.

**Applications of EVA to cargo transport by motor proteins.**

Motor protein is a general term for proteins that move and function using energy obtained from adenosine triphosphate (ATP) hydrolysis. Here, we focus on kinesin and dynein among such motor proteins, which are responsible for cargo transport and form the basis of intracellular material logistics (Fig. 2a). Particularly, we investigate anterograde transport by kinesin and retrograde transport by dynein in the axons of neurons. In the logistics of neurons, both anterograde transport, which carries synaptic materials to the terminals, and retrograde transport, which collects waste materials, are important. Because logistics in long axons is particularly important for neurons, and disruptions in this process are often associated with neurological diseases (Guedes-Dias, Holzbaur 2019; Keefe et al. 2023), the application of EVA to axonal transport is considered to be significant to help elucidate disease mechanisms in the future by increasing *in vivo* physical quantities we can estimate.

Cargo motion can be observed using fluorescence microscopy (Fig. 2a) (Hayashi et al. 2018; Naoi et al. 2024). Then the transport velocity is measured by analyzing the recorded movies obtained from the fluorescence microscopy. Note that fluorescence observation is a standard technique in biophysics, and this review will not discuss the methods of fluorescence observation. In *in vivo* cases, the measured velocity values varied affected by the size of a cargo and the number of motor proteins involved in the cargo transport (Fig. 3a). Especially since cargo sizes can differ by a factor of 10, the average velocity value reflects the cargo size rather than the performance of the motor proteins due to high viscosity in cells, unlike the situations in *in vitro* single-molecule experiments. The differences in the mechanical properties of the two different motor proteins, kinesin and dynein, are also less likely to be reflected in the velocity values, because the influence of cargo size on the velocity values is larger than the differences in the two motors. Then, the difference and true performance of the motor proteins can be considered to appear in the limit of small cargo sizes. Under this condition, it is believed that the difference and true performance of motor proteins can be accurately evaluated using EVA. In this review, in the following, we introduce the application of EVA to axonal transport in two different species: *C. elegans* worms and mice.

## *C. elegans* worms

The measurement method of transport velocity using fluorescence imaging in living *C. elegans* worms, was explained in the original paper (Naoi et al. 2024), noting that fluorescent proteins were labeled on the cargo for the fluorescence microscopy (Fig. 2a). Approximately 2,000 velocity data were collected from about 200 worms for anterograde and retrograde transport. One thing to be careful about is the fact that a measurement difficulty in live worms caused the small size of $M$ ($M = 10$ for this experiment), which is typically set to of the order of 1000 to obtain a correct extreme value distribution, highlighting a key problem in the application of EVA to nanoscale biological systems, which is discussed in the Discussion section.

Using the block maximum dataset $\{v_{\max}^i\}$ ($i = 1, \cdots, n$ where $n$=228 for anterograde transport and $n$=217 for retrograde transport), the return-level plot was calculated (it was calculated concretely by using the ismev and evd packages in R (R Core Team 2018)) (Fig. 2b). $\xi < 0$ for anterograde transport and $\xi \sim 0$ for retrograde transport. For anterograde transport, the return level plot shows a convergent behavior as $r_p$ becomes larger, a property specific to a Weibull distribution ($\xi < 0$), and the extreme value $V_{\text{ex}}$ was estimated to be 4.0±0.4 μm/s using Eq. (2).

Because the return-level plot of the retrograde transport (Fig. 2b) shows $\xi \sim 0$, and $V_{\text{ex}}$ cannot be estimated from Eq. (2) for retrograde transport by dynein, noting that the finite maximum exists only for the case $\xi < 0$. In the following simulation section, we explore why dynein does not exhibit a behavior of Weibull distribution.

## Mice

To investigate a mammalian case, EVA was also applied to examine the velocities of synaptic cargo transport by motor proteins in mouse hippocampal neurons (Fig. 2c, left), as originally reported in our previous study (Hayashi et al 2021). The return-level plot with $\xi < 0$ was observed for anterograde transport. Then the maximum velocity for anterograde transport in mice hippocampal neurons was calculated to be 5.6±1.4 μm/s using Eq. (2). The maximum velocity was higher than that of the worms. Comparing the maximum velocities among various species is a future important issue. Return-level plot with $\xi > 0$ was also observed for retrograde transport (Fig. 2c, right). It is known that dynein exhibits different mechanical properties depending on the species, but it is expected to have similar mechanical properties in both worms and mice, judging from the similar behaviors of the return level plots.

**Numerical simulation**

In our previous paper on the application of EVA to motor proteins (Naoi et al. 2024), the opposite signs of $\xi$ shown in the return level plots (Fig. 2b, c) was attributed to the different force (load)-velocity relationship between kinesin and dynein. The difference in the convexity of force (load)-velocity relationships (Fig. 3b), which have been clarified based on *in-vitro* single-molecule studies using optical tweezers, reflects the different mechanisms underlying the walking behavior of motor proteins on microtubules, although the biological meanings of convexity are not clearly understood yet. Due to high viscosity in cells, a motor protein experiences a load proportional to the cargo size transported by it. Therefore, we predicted that the behavior of velocity values *in vivo* would be influenced by the force (load)-velocity relationship. The observed extreme velocity values, which are close to those under the zero-load condition, are particularly related to the properties of the force (load)-velocity relationship under low load condition areas (green and yellow area depicted in Fig. 3b, right). The property that $\xi \sim 0$ for the retrograde velocity data was attributed to the fact that the force (load)-velocity relationship was concave. The steep velocity decrease for the case of the concave force (load)-velocity relationship in the low-load condition (green area in Fig. 3b) caused a major variation in the larger velocity values, and this behavior tends to generate a major variation in velocity. Then large gaps generated between $v_{\text{sim,max}}^{i}$ and $v_{\text{sim,max}}^{i+1}$ for a large $i$ in this case. This gap caused the non-convergence of $v_{\max}^{i}$ and $\xi \sim 0$ as a result. In the first place, the occurrence of small cargos, which are thought to generate high velocities, is a rare event. As a result, in a concave force (load)-velocity relationship, large velocity values are less likely to converge due to this rarity.

We can actually reproduce the behavior observed for anterograde and retrograde transport using a theoretical model of the force (load)-velocity relationship derived based on the mechanisms underlying ATP hydrolysis by motor proteins (Sasaki et al. 2018). The force (load)-velocity relationship for the model is represented as follows:

$$v_{\text{three}}(F) = (k_{01} - k_{02})l$$

$$\frac{1}{k_{01}} = \frac{1}{\kappa_1} + \frac{1}{\lambda_1} e^{d_1 F/k_B T}$$

$$\frac{1}{k_{02}} = \frac{1}{\lambda_2} e^{-d_2 F/k_B T} \qquad (3)$$

See reference (Sasaki et al. 2018) for the definitions and values of parameters. The differences in the model parameters (Eq. (3)) for kinesin and dynein were resulted in the different convexities of the force-velocity relationship. Fig. 3c (top) represents the $\{v_{\text{sim}}^{i}\}$ (4000 data) obtained from the simulations using the model (Eq. (3)). $\{v_{\text{sim,max}}^{i}\}$ were chosen from $\{v_{\text{sim}}^{i}\}$ ($M = 10$). Using these $\{v_{\text{sim,max}}^{i}\}$ ($n = 400$), we calculated the return-level plots for kinesin and dynein (Fig. 3c, bottom). We found that the tendency that $z_p$ ($= v_{\text{sim,max}}^{i}$ arranged in ascending order) did not converge for a large $r_p$ (Roughly, $r_p$ is the sample number ($1 \leq r_p \leq 400$)) in the case of dynein, *i.e.*, the gaps created between $v_{\text{sim,max}}^{i}$ and $v_{\text{sim,max}}^{i+1}$ for a large $i$. This is because a large velocity value is likely to be generated in the case of a concave force (load)-velocity relationship, owing to its steep slope in the load-sensitive regime.

**Discussion**

Typically, a block size (the number of elements in a block, $M$) of around 1000 is used to obtain a correct extreme value distribution (Coles 2001). In comparison, our study uses an extremely small $M$, around 10. This is due to the limited cargo transport observable in a single *C. elegans* worm, resulting in an $M$ value of around 10. It is difficult to collect a large number of samples within living organisms or cells. This is a fundamental issue in biological systems. In our research, we checked that while the parameters $\mu$, $\sigma$, and $\xi$ showed dependency on $M$ due to its small size, the qualitative results of $\xi < 0$ for anterograde and $\xi > 0$ for retrograde transport were independent of $M$ (Naoi et al. 2024). Due to the small sample size, it may be helpful to use other statistical methods, such as a bootstrap method, to cope with outliers of data when we apply EVA (Naoi et al. 2024).

There is one thing that should be mentioned regarding the conclusion obtained from the numerical simulation depicted in Fig. 3c. We attributed the property $\xi \sim 0$ in the extreme value data for retrograde transport (specifically the rare occurrence of large velocity values and the divergence observed in the return level plot at large $r_p$ values) to a concave force (load)-velocity relationship. However, several mechanisms could lead to the occasional large velocity values in retrograde transport, such as active non-equilibrium fluctuations originating from the cellular cytoskeleton (Chaubet et al. 2020; Mizuno et al. 2007) and the participation of multiple molecular motors (Fig. 3a) (Rai et al. 2013). In the future, we wish to further investigate the origin of large velocity values observed for retrograde transport.

To expand the application of EVA in nanoscale biological systems, it is crucial to consider applying EVA to systems entirely different from motor proteins. Inspired by the poster presentation by Takaramoto and Inoue at the IUPAB2024 Congress (Takaramoto and Inoue 2024), we became interested in the analysis of droplet sizes formed by liquid-liquid phase separation (LLPS). Within cells, there are membrane-less organelles formed through LLPS, such as stress granules and nucleoli. These organelles are formed through local concentration and separation caused by interactions between specific proteins and RNA, playing critical roles in cellular functions. While membrane-less organelles formed by LLPS are reversibly assembled and disassembled under normal conditions, in pathological states, this process can break down, leading to the formation of irreversible aggregates. In neurodegenerative diseases such as ALS and Parkinson's disease, abnormal irreversible aggregation of proteins has been reported to cause neuronal cell death. Takaramoto and Inoue investigated the mechanism of liquid condensate formation by the Parkinson's disease-related protein α-synuclein (αSyn). A previous study (Hoffmann et al. 2021) has shown that when αSyn coexists with the neuronal protein synapsin, αSyn is incorporated into the droplets formed by synapsin and can undergo phase separation at lower concentrations than when αSyn forms droplets alone. Takaramoto and Inoue studied the detailed molecular mechanism of droplet formation by αSyn in the presence of synapsin. Takaramoto and Inoue confirmed droplet formation in solution of αSyn and synapsin (*in vitro* experiment) (Fig. 1 of Reference (Takaramoto and Inoue 2024)), and obtained the droplet size distribution. From our preliminary application of EVA to the droplet size distribution, we found that this distribution follows the Fréchet type, which does not have a finite maximum value (unpublished data). On the other hand, within neurons, a maximum value for droplet size is expected to exist due to the fixed cell size. It is important to compare both the droplet formation in the *in vitro* experiment and cellular environments. In addition to experiments, molecular simulations have emerged as powerful tools to elucidate the physicochemical principles of biomolecular condensates, complementing experimental approaches. At Joseph's keynote session at IUPAB2024, we learned the potential of LLPS-specific simulations (Mpipi) (Joseph et al. 2021). Pursuing whether the maximum value of LLPS droplet size is related to the mechanism of disease onset and describing life sciences through extreme values rather than averages is a crucial task for the future.

**Declaration**
No funding was received for conducting this study.
**Conflict of interest**
The authors declare no competing interests.
**Data availability**
Data sharing is not applicable to this article as no new data were created.

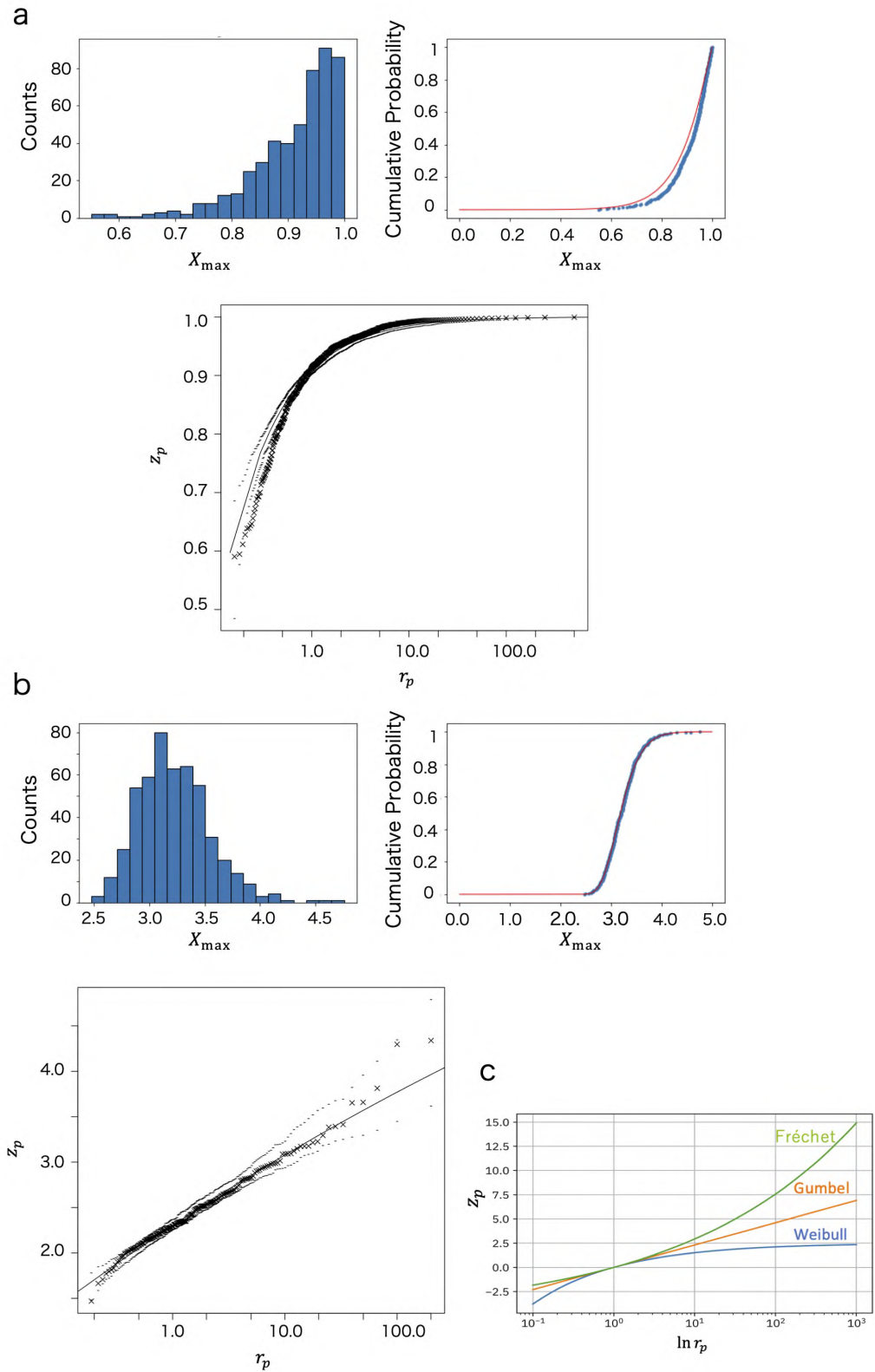

Fig. 1 Simulation using random numbers. The cases for uniform random numbers ranging from 0 to 1 ($M = 10$, $n = 500$) (**a**) and Gaussian random numbers ($M = 1000$, $n = 500$) (**b**). Each panel represents the counts, cumulative probability and return level plot for the block maximum. The shape parameter $\xi < 0$ for the uniform random numbers, and $\xi \sim 0$ for the Gaussian random numbers. Red lines represent $G(X_{\max})$ (Eq. (1)). (**c**) Examples of extreme value distributions.

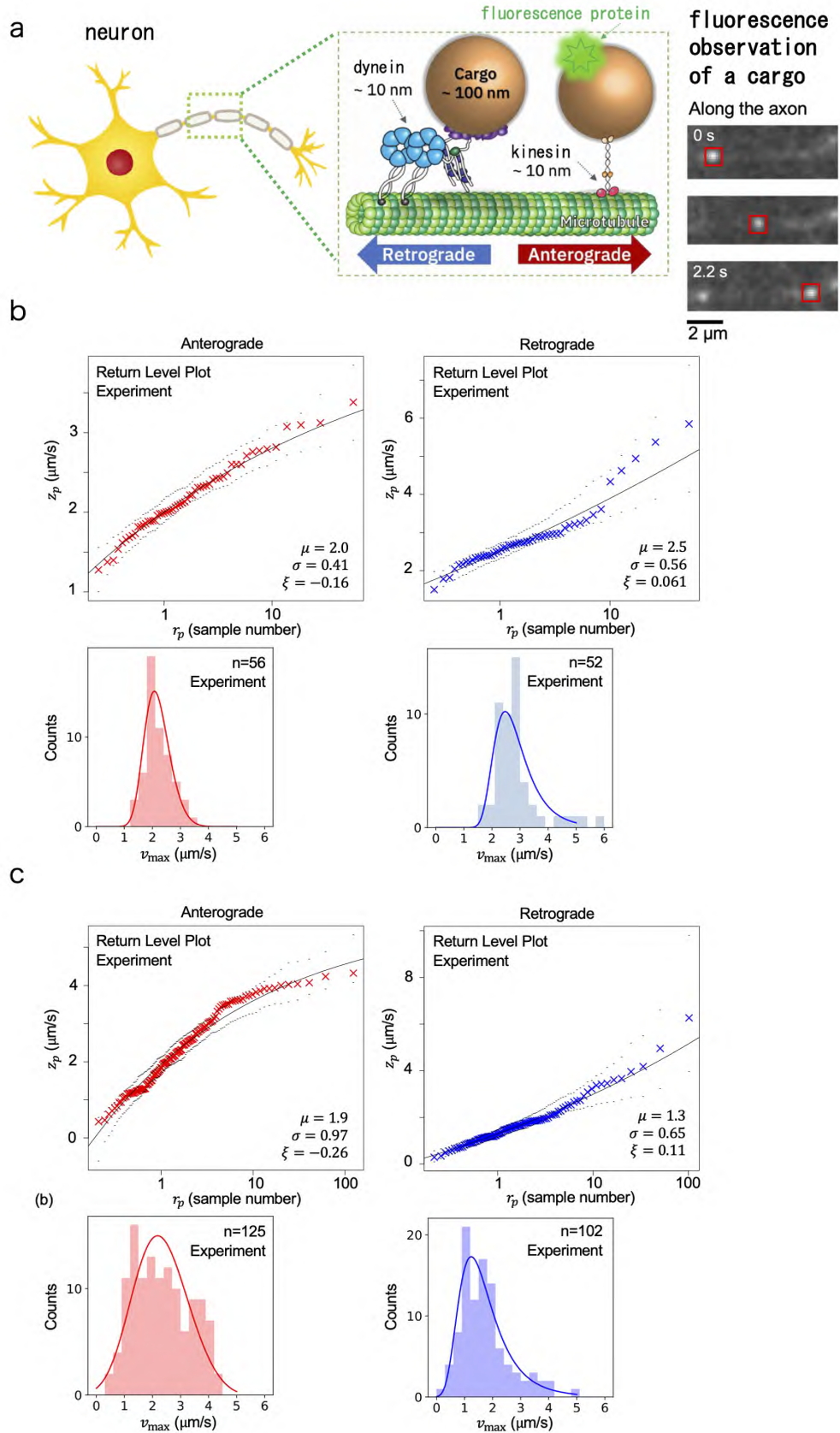

Fig. 2 Schematics of synaptic cargo transport by motor proteins in the axon of a neuron. **a** A cargo is anterogradely transported by KIF1A (kinesin) and retrogradely by cytoplasmic dynein. **b**, **c** Return level plot and histogram of the block maximum velocity data of synaptic cargo transport by motor proteins. Results for anterograde (red) and retrograde (blue) transport, in the cases of the motor neurons of *C. elegans* worms (Naoi et al. 2024) (**b**), and the hippocampal neurons of mice (Hayashi et al. 2021) (**c**).

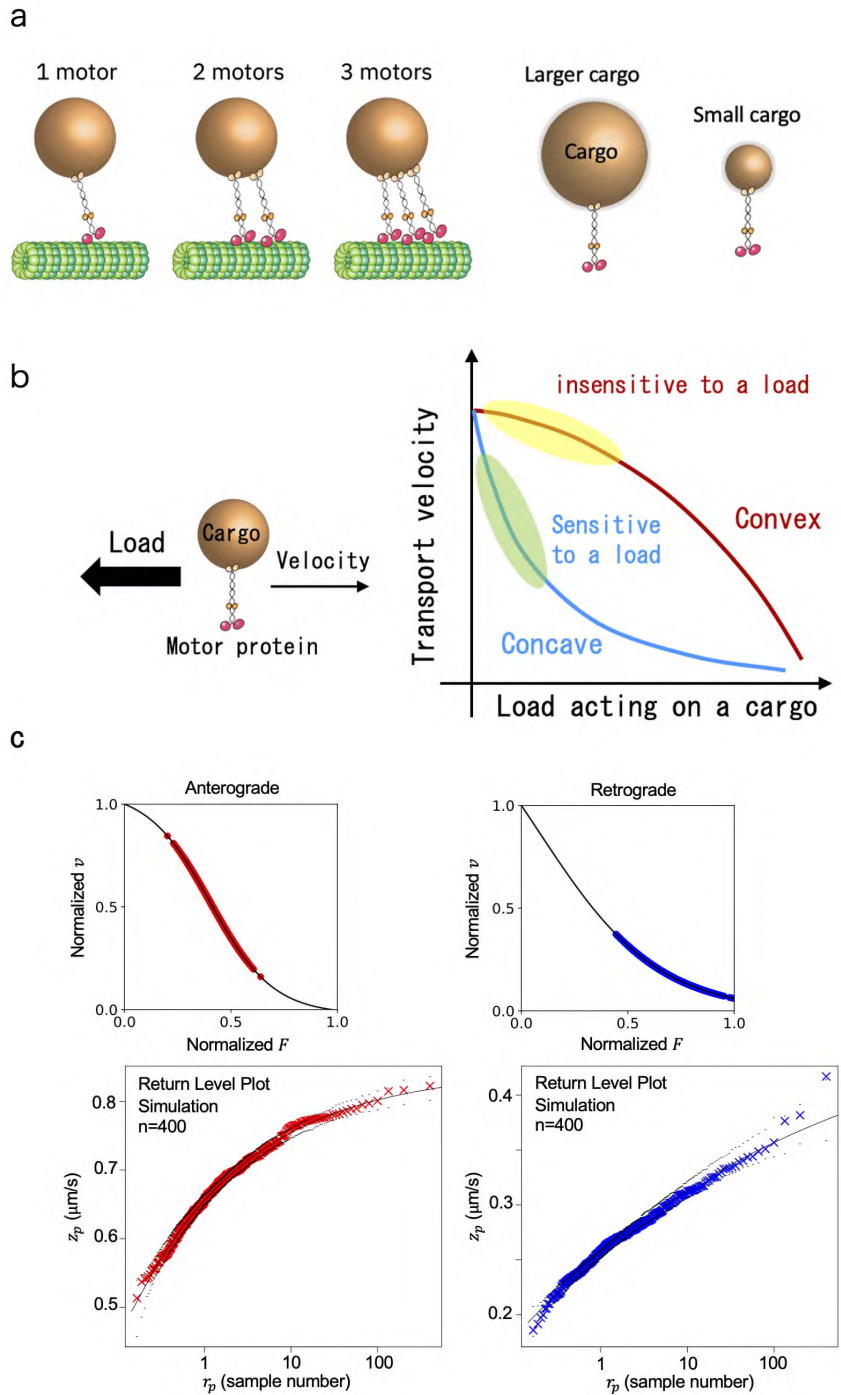

Fig. 3 Physical parameters of intracellular cargo transport by motor proteins. **a** The variety in the values of cargo transport resulted from the number of motor proteins and cargo sizes. **b** Load dependence of velocity in convex and concave cases. An existence of a cargo can be a load in the axons of neurons, because of high viscosity in the cells. **c** Simulation results for the model (Eq. (3)).